\documentclass[final]{siamltex}

\usepackage{graphicx}
\usepackage{epsfig}
\usepackage{amsmath,amssymb}
\def\(({\left(}
\def\)){\right)}

\def\[[{\left[}
\def\]]{\right]}

\newcommand{\be}{\begin{equation}}
\newcommand{\ee}{\end{equation}}
\newcommand{\bea}{\begin{eqnarray}}
\newcommand{\eea}{\end{eqnarray}}

\title{Quiet Planting in the Locked Constraint Satisfaction Problems}

\author{Lenka Zdeborov\'a\thanks{Theoretical Division and Center for Nonlinear Studies, Los Alamos
  National Laboratory, NM 87545 USA ({\tt lenka@lanl.org}).}
        \and Florent Krzakala\thanks{CNRS and ESPCI ParisTech, 10 rue Vauquelin, UMR 7083
  Gulliver, Paris 75000 France ({\tt fk@espci.fr}). Theoretical Division and Center for Nonlinear Studies, Los Alamos
  National Laboratory, NM 87545 USA.}}

\begin{document}

\maketitle

\begin{abstract}
  We study the planted ensemble of locked constraint satisfaction
  problems. We describe the connection between the random and planted
  ensembles.  The use of the cavity method is combined with arguments
  from reconstruction on trees and the first and the second moment
  considerations. Our main result is the location of the hard region in the planted ensemble. In a part of that hard region instances have
  with high probability a single satisfying assignment.
\end{abstract}



\pagestyle{myheadings}
\thispagestyle{plain}

\markboth{L. ZDEBOROV\'A AND F. KRZAKALA}{QUIET PLANTING IN THE LOCKED
  CONSTRAINT SATISFACTION PROBLEMS}

Constraint Satisfaction Problems (CSPs) are very general in their
nature: Consider a set of $N$ discrete variables and a set of $M$
Boolean constraints; the problem consists in finding a configuration
of variables that satisfies all the constraints or in proving that no
such configuration exists. As such, CSPs are a subject of interest in
many different fields such as computer science, discrete mathematics,
physics, engineering and computational biology. Random ensembles of
CSPs are a fertile source of research activity; as hard benchmarks
they serve for testing new algorithmic ideas
\cite{CheesemanKanefsky91,MitchellSelman92}, they are used to create
efficient coding schemes \cite{Gallager62,Gallager68}, to model
complex glass forming liquids \cite{BiroliMezard02,KrzakalaTarzia08},
or to understand the origin of average computational hardness
\cite{MezardParisi02,ZdeborovaMezard08}. Combining know-how from many
branches of mathematics, computer science and statistical physics
seems to be a fruitful strategy for the understanding of these
stunning objects and of their very rich behavior.

The most commonly studied random ensembles of CSPs are created by
choosing the graph of variables and constraints as a random bipartite
graph with a certain left and right degree distribution. Another
natural way of creating a random instance, called planting, is to
first assign a configuration to all variables, and then to choose only
constraints compatible with this configuration. Both these ensembles
can be useful to mimic instances created in some practical
applications, as e.g. the low density parity check codes
\cite{Gallager62}. In particular planted instances may be created in
adaptive situations when only constraints satisfied by the current
state of variables can be added.

By planting, we create by definition a satisfiable instance. Such
instances are particularly useful as benchmarks to evaluate the
performance of incomplete solvers, such as stochastic local search
\cite{SelmanKautz96}. Based on the example of the planted
K-satisfiability problem it is, however, often anticipated that the
planted ensemble is algorithmically easier than the random one because
a bias towards the planted assignment is created in the graph. Also,
for most of the studied problems, it was proven that at large density
of constraints is it indeed easy to find a satisfying assignment near
to the planted one, see
e.g. \cite{Ben-ShimonVilenchik07,Coja-OghlanKrivelevich07,FeigeMossel06}. 
On the other hand if the planted ensemble would be algorithmically
hard in some region of parameters than these instances could serve
as one-way functions and have applications in cryptography. Yet,
compared to the random ensemble, relatively little is known about the
existence, size and properties of algorithmically hard regions in the
planted ensemble.

In this paper we study a way of planting an assignment which changes
only in a minimal way the properties of the random ensemble. We call
this a {\it quiet planting}. Although the concept of {\it quiet planting}
was introduced in \cite{KrzakalaZdeborova09}, many result were
actually first demonstrated and used as a tool for proofs in
\cite{AchlioptasCoja-Oghlan08}. Both these works, however, were mainly
concentrated on the coloring problem (and on hyper-graph
bi-coloring). In the present work we generalize this idea and we will
focus on quiet planting in the so-called locked CSPs, introduced
recently in \cite{ZdeborovaMezard08,ZdeborovaMezard08b}. On one hand,
the locked CSPs have a very interesting phase diagram which
description-wise is much simpler than the one of the graph coloring or
K-satisfiability. On the other hand they are much harder
algorithmically and the boundaries between the easy and hard regions
are, unlike in the coloring or K-satisfiability, relatively well
understood (at least on the heuristic level of the cavity method
\cite{ZdeborovaMezard08,ZdeborovaMezard08b}). This special behavior
stems from the fact that in the locked problems the space of solutions
consists of points separated by an extensive (i.e. $\Theta(N)$)
Hamming distance instead of clusters of solutions.

Here, we combine the idea of quiet planting with the special behavior
of the locked CSPs and obtain random CSPs ensembles with many
interesting properties that are summarized in Sec.~\ref{sec:sum} in
the context of related works. In Sec.~\ref{sec:def} we introduce the
necessary definitions and notations, and in Sec.~\ref{sec:loc} we
summarize the phase diagram of the locked problems derived in
\cite{ZdeborovaMezard08,ZdeborovaMezard08b}. In Sec.~\ref{sec:pl} we
argue about the equivalence between the random and the planted ensemble
based on heuristic considerations and on a second moment computation.  In
Sec.~\ref{sec:sin} we describe the interesting phase where instances
of our problems have with large probability a single
solution. Finally, in Sec.~\ref{sec:alg} we discuss the algorithmic
hardness of the planted instances, and compute the critical degree
beyond which instances become easy to solve. We conclude the paper with
a list of open problems in Sec.~\ref{con}.

\section{Main results and related works}
\label{sec:sum}

The results of this paper apply to the {\it factorized locked} CSPs,
see Defs. \ref{def:loc}, \ref{def:fac}. We list in six points the
most important contributions of the present article:

\begin{itemize}
\item[(i)] {\bf The planted configuration is equivalent to randomly
    sampled satisfying assignment:} The idea of {\it quiet} planting
  is to plant a configuration that in many important properties does
  not differ from a satisfying assignment sampled uniformly at random
  on the resulting graph. Such a problem is closely related to the
  reconstruction on trees
  \cite{EvansKenyon00,Mossel01,MezardMontanari06} where we consider an
  assignment taken uniformly at random from all the satisfying
  ones. Constructing such an assignment on a tree is always possible
  in polynomial time, as the exact marginal probabilities can be
  obtained via the belief propagation (BP) algorithm. On random graphs,
  uniform sampling is in general exponentially costly. However, the
  quiet planting can be achieved asymptotically on the factorized
  CSPs, see Def.~\ref{def:fac}. In a statistical physics language
  quiet planting is possible whenever the quenched entropy equals the
  annealed one, see condition (\ref{annealed}). The possibility of
  quiet planting and its relation to a concentration property
  (\ref{annealed}) was previously discussed for other constraint
  satisfaction problems in
  \cite{MontanariSemerjian06,AchlioptasCoja-Oghlan08,KrzakalaZdeborova09}.
\item[(ii)] {\bf Equivalence between the planted and the random ensemble in
    the satisfiable phase:} Many properties of the planted ensemble
  created via quiet planting can be deduced from the properties of the
  purely random ensemble. Among others, in the satisfiable phase the
  random and planted ensembles are asymptotically equivalent, see
  Def.~\ref{def:equ}. Such equivalence can also be established
  rigorously based on a second moment argument, as in
  \cite{AchlioptasCoja-Oghlan08}. In the factorized locked problems treated in this article the second moment is able to pin the
  satisfiability threshold sharply and hence the equivalence between the planted and the random ensemble holds in the whole satisfiable phase.
\item[(iii)] {\bf Equivalence between planted and satisfiable
    ensembles:} Based on heuristic (cavity) arguments we conjecture
  that the planted ensemble is in the factorized locked problems asymptotically
  equivalent to the ensemble of satisfiable instances in the whole range of $\Theta(1)$ constraint densities. In particular,
  this means that one can recognize easily almost all the rare
  satisfiable instances above the robust reconstruction threshold by
  using belief propagation. We stress here that we do not expect this
  equivalence to be true in the non-locked CSPs, such as graph
  coloring or K-satisfiability.
\item[(iv)] {\bf Easy-hard-easy transition:} Next to the interesting
  conceptual results, the most important practical result is the
  identification and the location of the region where the
  instances from the planted ensemble of the factorized locked
  problems are on average computationally hard. We show that an easy-hard-easy
  pattern for finding a solution appears in the planted ensemble as
  the constraint density is increased, where {\it easy} means that an
  on-average polynomial algorithm is known, while {\it hard} means
  that no on-average polynomial algorithm is known (and maybe does not
  exist). We conjecture that the two boundaries of the hard phase
  correspond to two different reconstruction thresholds -- the onset
  of hardness coincides with the {\it small noise} reconstruction
  threshold \cite{ZdeborovaMezard08b}, called the dynamical transition
  in the physics literature \cite{KrzakalaMontanari06}, and the end of
  the hard region is given by the threshold for the {\it robust}
  reconstruction \cite{JansonMossel04}. This last point also
  corresponds to the Kesten-Stigum bound for the canonical
  reconstruction on trees \cite{KestenStigum66,KestenStigum66b}, and to
  the spin glass local instability in the purely random ensemble
  \cite{MezardParisi87b}. We also show that outside the hard region
  algorithms based on belief propagation are able to find solutions
  efficiently. In particular in the high average degree easy region,
  the belief propagation algorithm converges directly to the planted
  solution.
\item[(v)] {\bf Hard satisfiable benchmarks:} Given we have located
  the values of parameters where the instances of the planted ensemble
  are hard, these can serve as very challenging satisfiable
  benchmarks. Such benchmarks are in particular interesting for the
  evaluation of regions in which incomplete solvers work ---or do not
  work--- in polynomial (linear) time.  Note that for complete
  exhaustive solvers the locked problems are not necessarily harder
  than the canonical K-satisfiability. On the contrary, locked constraints
  produce more implications when a variable is fixed, hence exhaustive
  branch-and-bound techniques might come to a decision relatively faster than in the
  random K-satisfiability.
\item[(vi)] {\bf Instances with unique satisfying assignment (USA):} We show that beyond the threshold corresponding to the satisfiability threshold in the random ensemble the planted instances have with high probability a
  single satisfying assignment (or a pair of them in case a global
  symmetry is present). Moreover depending on the constraint density
  these USA instances can be found in
  the hard or in the easy region.  Some USA instances are extensively
  used in evaluation of quantum algorithms, see
  e.g. \cite{YoungKnysh08,FarhiGoldstone01}. In these previous works
  these instances are, however, generated with exponential cost, and
  their classical typical computational hardness has not been evaluated. 
\end{itemize}

A large part of our results is based on the heuristic cavity method approach \cite{MezardParisi01}. We were also able to prove part of our results for the $R$-in-$K$ SAT problem on random regular graphs using computations of the second moment and the expander property. This includes some results about the equivalence between the planted and random ensembles in the satisfiable phase, and the uniqueness of the satisfying assignment in the unsatisfiable phase. Completing and extending these proofs to the other locked factorized CSPs should be possible although more involved.

\begin{table}[!ht]
  \caption{Sketchy summary of the properties of the different phases in the random ensemble of the factorized locked problems, the parameter $l$ is the average number of constraints in which a variable appears. The three thresholds $l_d$, $l_s$ and $l_l$ are defined in detail later in the paper.}
\begin{center}
  \begin{tabular}{|l||c|c|c|c|} \hline 
RANDOM & $l < l_d$ & $l_d < l < l_s$ & $l_s < l < l_l$ & $l_l <l $ \\ \hline \hline 
BP, random init. & converges & converges & converges & \, does not \\ \hline 
BP fixed point & uniform & uniform & uniform & $\times$ \\ \hline \hline
$\#$ of solutions & exponential & exponential & none & none \\ \hline
finding solution \hspace{12mm}& easy & hard & $\times$ & $\times$ \\ \hline \hline 
reconstruction & not possible & possible & $\times$ & $\times$ \\ \hline
\end{tabular}  
\end{center}
\label{tab:summary_ran}
\end{table}

\begin{table}[!ht]
  \caption{The same as Tab.~\ref{tab:summary_ran} for the random planted ensemble, its definition is given in Sec.~\ref{sec:pl}.}
\begin{center}
  \begin{tabular}{|l||c|c|c|c|} \hline PLANTED & $l < l_d$ & $l_d < l <
    l_s$ & $l_s <l < l_l$ & $l_l < l $ \\ \hline \hline BP, random
    init. & converges & converges & converges & converges \\ \hline
    BP fixed point & uniform & uniform & uniform & planted \\ \hline
    \hline BP, planted init.  & converges & converges & converges &
    converges \\ \hline BP fixed point & uniform & planted & planted &
    planted \\ \hline \hline $\#$ of solutions & exponential & exponential
    & one/two & one/two \\ \hline finding solution & easy & hard &
    hard & easy \\ \hline \hline reconstruction & not possible &
    possible & possible & possible \\ \hline robust recons. & not
    possible & not possible & not possible & possible \\ \hline
\end{tabular}  
\end{center}
\label{tab:summary_pl}
\end{table}

\section{Definitions and notations}
\label{sec:def}

In this section we specify the class of constraint satisfaction
problems to which our results apply. The crucial notions will be the
definition of a {\it locked} \cite{ZdeborovaMezard08} and {\it
  factorized} constraint satisfaction problem. It is only on the
factorized problems where there is a very close relation between the
usual random and the planted ensemble, as discussed in
\cite{KrzakalaZdeborova09}. It is also the fact that in the locked
problems solutions (i.e. satisfying assignments) are mutually far from each other in terms of their
Hamming distance \cite{ZdeborovaMezard08} that makes them particularly
interesting for considerations in this context.

\begin{definition}
  A {\em constraint} $a$ containing $K$ variables, the domain of each
  variable being $X$, is a function from $X^K$ to $\{0,1\}$. If the
  function evaluates to $1$ ($0$) we say that constraint $a$ is
  satisfied (not satisfied). A constraint is {\em locked} if and only
  if there are no two satisfying assignments of variables which would
  differ in a single value (out of the $K$ ones).
\end{definition}
In this paper we will consider for concreteness binary variables,
that is $X=\{0,1\}$, but the results are generalizable to larger domain
sizes.
\begin{definition}
  \label{def:loc}
  A {\em constraint satisfaction problem} consists in deciding if
  there exists a configuration of $N$ variables which satisfies
  simultaneously a set of $M$ constraints. A constraint satisfaction
  problem is called {\em locked} if and only if all the $M$ constraint
  are locked and each of the $N$ variables belongs to at least two
  different constraints.
\end{definition}
Thus anytime we speak about a locked problem we implicitly suppose
that the corresponding factor-graph does not have any leaves
(variables of degree one). The degree of a variable is defined as the
number of constraints to which the variable belongs, while the degree
of a constraint is the number of variables it contains.

We shall illustrate our findings on the so called occupation
constraint satisfaction problems \cite{Mora07,ZdeborovaMezard08b}.
\begin{definition}
  In {\em occupation problems} every constraint $a$ depends only on
  the sum of the variables it contains. Thus every occupation constraint
  containing $K_a$ variables can be characterized by a binary $K_a+1$
  component vector $A_a$ such that the constraint is satisfied if and
  only if the sum $r$ of the $K_a$ variables is such that $A_a(r)=1$.
\end{definition}

An occupation constraint $a$ is locked if and only if for all
$i=0,\dots,K_a-1$ we have $A_a(i)A_a({i+1})=0$. We will consider
occupation problems where every constraint contains $K$ variables and
is given by the same vector $A$. To give an example of this notation,
the vector $A=0100$ corresponds to the $1$-in-$3$ SAT problem (also
called exact cover), which is indeed locked. The vector $A=0110$
corresponds to the hyper-graph bi-coloring problem, which is not
locked (since there are two neighboring $1$s). Many other examples can
be found in \cite{ZdeborovaMezard08b}. For problems which do not have
other name established in the literature, we will use the notation
i-or-j-...-in-K SAT for a vector $A$ with non-zero components
$A(i)$,$A(j)$, etc.

\begin{figure}[!ht]
  \begin{center}
  \resizebox{3cm}{!}{\includegraphics{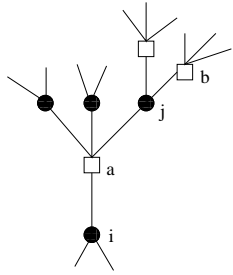}}
  \end{center}
  \caption{\label{fig_cav} Part of the factor graph to illustrate the
    meaning of indices in the belief propagation equations
    (\ref{BP1}).}
\end{figure}

Let us now write the belief propagation (BP) equations
\cite{Pearl82,KschischangFrey01,MezardMontanari07} for the occupation
constraint satisfaction problems. The basic quantities in BP are
messages. We define $\mu_{s_i}^{a\to i}$ as the probability (over all
satisfying assignments) that the variable $i$ has value $s_i$ given
that $i$ belong only to constraint $a$. The BP equations approximate
these probabilities $\mu_{s_i}^{a\to i}$ by messages $\psi_{s_i}^{a\to
  i}$ by assuming that the factor graph \cite{KschischangFrey01}
underlying the CSP is a tree \be \psi_{s_i}^{a\to i} =
\frac{1}{Z^{a\to i}} \sum_{\{s_j\}} \delta_{A({s_i+\sum_{j}
    s_j}),1}\prod_{j\in
  \partial a-i} \prod_{b\in \partial j-a} \psi_{s_j}^{b\to j} \,
, \label{BP1} \ee where $Z^{a\to i}$ is a normalization constant
assuring $\psi_1^{a\to i}+\psi_0^{a\to i}=1$, $\partial a$ is the set
of neighbors of $a$, $\partial a-i$ are neighbors of $a$ except $i$,
and the sum over $\{s_j\}$ is over all values variables $s_j$ can
take. Fig.~\ref{fig_cav} shows the corresponding part of the factor
graph.

We define $\nu^i$ to be the probability (over all satisfying
assignments) that a variable $i$ is occupied. The BP estimate of the
probability that a variables $i$ is occupied is \be \chi^{i} = \frac{
  \prod_{a\in \partial i} \psi_{1}^{a\to i}}{ \prod_{a\in \partial i}
  \psi_{1}^{a\to i}+ \prod_{a\in \partial i} \psi_{0}^{a\to i}}\,
, \label{marginal} \ee Note that if an assignment $\{\sigma\}$ is a
solution of the locked problem, then $\psi_{\sigma_i}^{a\to i}=1$,
$\psi_{\neg \sigma_i}^{a\to i}=0$ is a fixed point of the BP equations
(\ref{BP1}). If the underlying factor-graph is a tree then the fixed
point of the BP equations is unique and $\mu_{s_i}^{a\to
  i}=\psi_{s_i}^{a\to i}$ and $\nu^i=\chi^{i}$ in the fixed point
(note that on a tree the problem is not locked). On a graph with
cycles this is not the case in general. We will call a BP fixed point
asymptotically ($N \to \infty$) exact on a random ensemble of graphs
if $\mu_{s_i}^{a\to i}=\psi_{s_i}^{a\to i} + o(1)$ and $\nu^i=\chi^{i}
+ o(1)$ for almost all $a$ and $i$ with high probability, where $N$ is the number of nodes in the
graph.

All our results are restricted to ensembles of constraint satisfaction
problems where at least one BP fixed point is asymptotically exact.
We define such ensembles in the remaining of this section.  A graph
ensemble is locally tree-like if the shortest loop going trough a
random node has w.h.p. length diverging as $N\to \infty$.  Families of sparse
random graphs, i.e. the degree distribution of variables $Q(l)$ does
not depend on $N$, are locally tree-like as long as the mean of
$Q(l)$ is finite. The variable degree distributions we will
be using mostly are:
\begin{itemize}
    \item Regular $Q(l)=\delta_{L,l}$.
    \item Truncated Poisson $Q(0)=Q(1)=0$, $Q(l)=c^l/[(e^c-1-c)l!]$
      for $l\ge 3$. The average degree in this case is $\overline l =
      c (1-e^{-c}) / [1-(1+c)e^{-c}]$.
\end{itemize}
In order to generate random graphs with a given variable degree
distribution, one can apply the following algorithm:
\begin{itemize}
   \item{Repeat until $KM=\sum_{i=1}^N l_i$: Draw $N$ random numbers $l_i$ from distribution $Q(l)$.}
   \item{Consider $K$ legs going from every constraint, order them arbitrarily and index them from $1$ to $KM$, consider $l_i$ legs from every variable $i$, order them arbitrarily and index them from $1$ to $KM$.}
   \item{Repeat until there are no double edges: Draw a random
       permutation $\pi$ of $KM$ numbers and connect $i$-th leg from
       constraints with $\pi(i)$-th leg from variables.}
\end{itemize}

We define an iteration of the belief propagation algorithm as taking all the KM edges $ai$ in a random order and updating the message $\psi_{s_i}^{a\to i}$ according to eq. (\ref{BP1}).
\begin{definition}
\label{def:fac}
A given instance of a constraint satisfaction problem is {\em
  factorized} if and only if the belief propagation equations
initialized randomly converge almost surely (with probability
approaching one as the number of variables $N \to \infty$) to a
uniform fixed point, i.e., the value of $\psi^{a\to i}$ is the same for almost all edges $ai$. 
\end{definition}

Note that it is a non-trivial task to provably decide if a problem
satisfies this definition, and the answer depends on the degree
distribution. In practice we generate a large random instance of the
problem, initialize BP randomly and iterate. We observe that the
result (i.e. if the condition in def.~\ref{def:fac} is satisfied or
not) is the same on almost all large random instances. The condition
of def.~\ref{def:fac} can hence be checked computationally with a small computer-time effort.

\begin{definition}
\label{def:ann}
Let ${\cal N}_G$ be the number of satisfying assignments of an instance of the constraint satisfaction problem $G$. We define the annealed entropy $s_{\rm ann}$ to be
\be
         s_{\rm ann} = \lim_{N\to \infty} \frac 1 N \log{ \mathbb  E({\cal N}_G)} \, , \label{eq:ann}
\ee  
where the expectation is over the graph ensemble. The quenched entropy is defined as 
\be
          s_{\rm quen}  =\lim_{N\to \infty}  \frac 1 N \mathbb E[\log{({\cal N}_G+1)}] \, .  \label{eq:quen}
\ee
\end{definition}
\begin{definition}
\label{def:Bethe}
Let us also define the Bethe entropy \cite{KschischangFrey01,MezardMontanari07} that is associated to any BP fixed point as 
 \be s=\frac{1}{N}\sum_a
\log{(Z^{a})} - \frac{1}{N} \sum_i (l_i-1) \log{(Z^i)}\,
. \label{eq:entropy_l} \ee where
\begin{subequations}
\label{eq:Z_locked}
\bea
     Z^{a}&=& \sum_{\{s_i\}} \delta_{A_{\sum_i s_i},1} \prod_{i\in \partial a} \left( \prod_{b\in \partial i-a} \psi_{s_i}^{b\to i} \right)\, , \label{Za}\\
     Z^i&=&  \prod_{a\in \partial i} \psi_{0}^{a\to i}+ \prod_{a\in \partial i} \psi_{1}^{a\to i}\, . \label{Zi}
\eea 
\end{subequations}
\end{definition}

The following statement stands on the basis of the cavity method: If a BP fixed point is asymptotically exact (for a given random graph ensemble) then the Bethe entropy (\ref{eq:entropy_l}) is equal to the quenched entropy (\ref{eq:quen}).

The following result was obtained in \cite{Mora07} (section 5.1.2) for the occupation constraint satisfaction problems, and we conjecture it is general: The annealed entropy (\ref{eq:ann}) is equal to the Bethe entropy evaluated in the uniform BP fixed point (when more than one uniform BP fixed point exists then consider the maximum of the Bethe entropy over the uniform fixed points). 

If the problem is factorized and the uniform BP fixed point is asymptotically exact then the annealed entropy is equal to the quenched entropy, $s_{\rm ann}=s_{\rm quen}$. Our results in the remaining of this paper apply to random constraint satisfaction problems where indeed $s_{\rm ann}=s_{\rm quen}$ (at least in some region of constraint densities). This condition is sometimes amenable to a rigorous proof, as it is in general weaker than $\mathbb E({\cal N}_G^2)<C\, [\mathbb E({\cal N}_G)]^2$, see \cite{AchlioptasCoja-Oghlan08}. If we are, however, interested in a fast heuristic check, then checking if the random constraint satisfaction problem is factorized may be more suitable.

\section{Basic properties of the random factorized locked problems}
\label{sec:loc}

For the locked problems, a detailed empirical analysis was done in
\cite{ZdeborovaMezard08,ZdeborovaMezard08b}. In this section we
summarize the most relevant results (heuristically reasoned
conjectures) of those works. It was found that a locked CSP is factorized in (at least) the two following cases:
\begin{itemize}
\item[(a)] {\bf Any locked problem on random regular graphs}, that is
  when every variable is contained in $L$ constraints. On regular
  graphs, the uniform fixed point of the BP equations
  then satisfies \bea
  \psi_0&=& \frac{1}{Z}  \sum_{r=0}^{K-1}   \delta_{A(r),1} {K-1 \choose r} \psi_1^{(L-1)r} \, \psi_0^{(L-1)(K-1-r)}\, , \label{RS1}\\
  \psi_1&=& \frac{1}{Z} \sum_{r=0}^{K-1} \delta_{A(r+1),1} {K-1
    \choose r} \psi_1^{(L-1)r} \, \psi_0^{(L-1)(K-1-r)}\,
  , \label{RS2} \eea where $Z$ is the normalization. For the
  probability that a variable in occupied one has in this case in the
  $N\to \infty$ limit \be \chi= \frac{ \psi^L_{1}}{ \psi_{1}^{L}+
    \psi_{0}^{L}}\, , \label{marginal_reg} \ee Let us call $x_r$ the
  probability that a constraint contains $r$ occupied variables. Then
  \be x_r = \frac{{K\choose r} \delta_{A(r),1} \psi_1^{r(L-1)}
    \psi_0^{(K-r)(L-1)} }{\sum_{t=0}^K {K\choose t} \delta_{A(t),1}
    \psi_1^{t(L-1)} \psi_0^{(K-t)(L-1)}}\, . \label{prob_1} \ee
\item[(b)] {\bf The balanced locked problems}
  \cite{ZdeborovaMezard08b}, are problems where the vector $A$ is
  symmetric, $A(i)=A(K-i)$ for all $i=0,\dots,K$ and this 0-1 symmetry
  is not spontaneously broken (that is when a satisfying assignment
  chosen uniformly at random has the same number of $0$'s and $1$'s up
  to a $o(N)$ factor). Note that the absence of the symmetry breaking
  might depend on the degree distribution $Q(l)$.  In the balanced
  locked problems the uniform BP fixed point
  $\psi_1=\psi_0=\chi=1/2$. For the probability that a constraint
  contains $r$ occupied variables we have here \be x_r =
  \frac{{K\choose r} \delta_{A(r),1} }{\sum_{t=0}^K {K\choose t}
    \delta_{A(t),1}}\, . \label{prob_2} \ee
\end{itemize}  
A particularly simple case of (a) is the $R$-in-$K$ SAT where $1\le R
\le K/2$. If every variable has $L$ connections and every constraint
has to contain exactly $R$ occupied variables, then the number of
occupied variables is exactly $MR/L$, and thus $\nu=R/K$.

The authors of \cite{ZdeborovaMezard08,ZdeborovaMezard08b} conjectured
that when the $N\to \infty$ limit of the Bethe entropy for a locked
problem is positive, then the Bethe entropy is equal to the quenched entropy, and the BP fixed point reached from random initialization is asymptotically exact. If the Bethe entropy is negative then no satisfying assignment exists with high probability.

The Bethe entropy for all the balanced locked problems reads
\be 
s\left(\overline l\right) = \log{2} +
\frac{\overline l }{K} \log{\left[2^{-K} \sum_{r=0}^K \delta_{A(r),1}
    {K\choose r}\right]}\, , \ee 
where $\overline l$ is the average degree of a variable (as we speak only about locked problems, the degree distribution has to have a zero weight on variables of degree zero and one). 
For all the locked problems on random regular (degree fixed to $L$) graphs the entropy reads
\be
s(L)= \frac{L}{K} \log{\left[ \sum_{r=0}^K \delta_{A(r),1} {K \choose
      r} \psi_1^{(L-1)r} \psi_0^{(L-1)(K-r)}\right]} - (L-1)
\log{\left[\psi_{0}^{L}+ \psi_{1}^{L}\right]} \, ,\label{ent_reg} 
\ee 
where $\psi_1$,
$\psi_0$ is the fixed point of eqs.~(\ref{RS1}-\ref{RS2}). This entropy
simplifies further for the $R$-in-$K$ SAT on regular graphs (where the values of the $\psi$s obey the simple form discussed previously)
where we get an explicit formula 
\be 
s(L)= \frac{L}{K}\log{K \choose R} -
(L-1)\, H\left(\frac{R}{K}\right) \, ,
\label{ent_RinK} \ee where
$H(x)=-x\log{x}-(1-x)\log{(1-x)}$ is the entropy function.

The satisfiability transition $l_s$ is then defined by \be {\rm
  satisfiability}\, \, {\rm threshold} \, \, l_s: \quad s(l_s)=0 \ee
for the corresponding entropy function.  For $\overline l<l_s$ the
problem has almost surely exponentially many solutions (the exponent
being given by $s(\overline l)$) whereas for $\overline l>l_s$ the
problem almost surely does not have any solution.

The authors of \cite{ZdeborovaMezard08,ZdeborovaMezard08b} also argued
about the existence of a second phase transition in the locked
problems, $l_d<l_s$, traditionally called in the physics literature
the dynamical transition because of its connection to dynamics of
glasses \cite{MontanariSemerjian06}. This critical point separates a
region where for $\{\sigma\}$ being a typical satisfying assignment
the $\psi_{\sigma_i}^{a\to i}=1$, $\psi_{\neg \sigma_i}^{a\to i}=0$
defines a {\it stable} fixed point of the BP equations (\ref{BP1}),
from a region where this does not hold anymore.  In other words, if an
infinitesimal perturbation is introduced to these messages, the
iteration of (\ref{BP1}) goes back to the solution-related fixed
point for  $l>l_d$, but not for $l<l_d$.
The authors of \cite{ZdeborovaMezard08,ZdeborovaMezard08b} also
conjectured that for $\overline l>l_d$ a typical solution does not
have solutions up to an extensive (i.e. $\Theta(N)$) Hamming distance,
whereas for $\overline l<l_d$ there are other solutions at
sub-extensive (i.e. $o(N)$) Hamming distance. Let us call the phase
corresponding to $\overline l>l_d$ the separated phase, and the one
corresponding to $\overline l<l_d$ the non-separated phase.

For the locked problems on regular graphs, the following inequality
always holds: $2 < l_d < 3$. In other words at $L=2$ the system is
in the non-separated phase while for $L\ge 3$ the solutions are always
separated and the solution-corresponding fixed points are stable. 
For the balanced locked problems whenever the degree of every variable is larger or equal to three the system is in the phase where solutions are separated. When the fraction of variables of degree two is positive, $Q(2)>0$, then the expression for $l_d$ follows \cite{ZdeborovaMezard08b}:
 \be \frac{ l_d }{ Q(2)} = 2(K-1) - 2 \,
\frac{\sum_{r=1}^{K-2} r\, {K-1 \choose r}\, \delta_{A(r+1),1} \,
  \delta_{A(r),0} \, \delta_{A(r-1),0} }{\sum_{r=0}^{K-2}
  \delta_{A(r+1),1} {K-1\choose r}}\, . \label{eq_ld} \ee

There is a deep connection between this dynamical threshold $l_d$ and
the reconstruction problem  \cite{EvansKenyon00,Mossel01,MezardMontanari06}. In the reconstruction problem one creates
a tree with the same degree properties as the random graph. Then one
considers a satisfying assignment chosen uniformly at random from all
the possible ones. The reconstruction problem finally consists in
deciding whether this assignment on the leaves of the tree contains
some information about the value assigned to the root. In the locked
problems the value of the root is always uniquely implied by the
values of the leaves, as follows from the very definition of these
problems. However, if an infinitesimal noise is introduced on the
leaves then there is no information left if and only if $\overline l <
l_d$. This value $l_d$ was called the {\it small noise} reconstruction
threshold in \cite{ZdeborovaMezard08b}.

To summarize, the random locked factorized problems are in the
non-separated phase for $\overline l \le l_d$, which was shown to be
algorithmically easy in
\cite{ZdeborovaMezard08,ZdeborovaMezard08b}. For $l_d \le \overline l
\le l_s$ the space of solutions is separated and it is then hard to
find any solution. For $\overline l \ge l_s$ no solution exists anymore.

\begin{table}[!ht]
  \caption{ The critical values for all the balanced locked problems up to $K=8$ on the regular and truncated Poissonian ensembles. We remind here that the vector $A$ codes for what are the allowed sums of variables around a constraint. We consider only problems where $A(0)=A(K)=0$, that do not have a trivial {\it all true} or {\it all false} satisfying assignment. The integer value $L_s$ (resp. $L_l$) is defined as the first larger or equal to $l_s$ (resp. $l_l$), the stars denote that $L_s=l_s$ (resp. $L_l=l_l$). For definition of the threshold $l_l$ see Sec.~\ref{sec:alg}. The corresponding values of $c$ are the coefficients that in the truncated Poisson distribution correspond to the average degree $\overline l$. The sign '$\times$' means that the problem ceases to be balanced before the instability arises. }
\begin{center}
  \begin{tabular}{|l||c|c||l|l|l||l|l|l|} \hline A &\, $L_s $&\, $L_l$
    &\, $c_{d}$&\, $c_{s}$&\, $c_{l}$&\, $ l_d$&\, $ l_s$&\, $
    l_{l}$\\ \hline
    00100      &3 &4*    & 1.256     & 1.853    &  2.821   & 2.513     & 2.827  & 3.434    \\ \hline 
    0001000    &4 &6*    & 1.904     & 3.023    &  4.965   & 2.856     & 3.576  & 5.144    \\ \hline
    000010000  &5 &8*    & 2.337     & 3.942    &  6.994   & 3.116     & 4.276  & 7.039    \\ \hline 
    5-{\rm in}-10   &5 &10*    & 2.660     & 4.794    &  8.999   & 3.325     & 4.944  & 9.009    \\ \hline 
    6-{\rm in}-12   &6 &12*    & 2.918     & 5.455    &  11.00   & 3.502     & 5.586  & 11.00    \\ \hline  
    01010      &4*&$\infty$& 1.904   & 3.594    & $\infty$ & 2.856     & 4      & $\infty$ \\ \hline
    0101010    &6*&$\infty$& 2.660   & 5.903    &  $\infty$& 3.325     & 6      & $\infty$ \\ \hline
    010101010  &8*&$\infty$& 3.132   & 7.978    & $\infty$ & 3.654     & 8      &  $\infty$\\ \hline
    0010100    &6 &46*   & 2.561     & 5.349    &  45.00   & 3.260     & 5.489  & 45.00    \\ \hline
    000101000  &7 &29*   & 2.975     & 6.650    &  28.00   & 3.542     & 6.708  & 28.00    \\ \hline
    001010100  &8 &$>100$ & 3.110    & 7.797    &  $>100$  & 3.638     & 7.822  &   $>100$ \\ \hline
    010010010  &6 &{\rm x}& 2.173    & 4.896    &  {\rm x} & 3.014     & 5.083  &   {\rm x}\\ \hline
\end{tabular}  
\end{center}
\label{tab:balanced}
\end{table}

\begin{table}[!ht]
  \caption{ The critical values for all the regular (non-balanced) locked problems up to $K=6$. The integer value $L_s$ (resp. $L_l$) is defined as the first larger or equal to $l_s$ (resp. $l_l$), the stars denote that $L_s=l_s$ (resp. $L_l=l_l$).}
\begin{center}
  \begin{tabular}{|l||c|c|} \hline A &\, $L_s $&\, $L_l$\\ \hline 0100 
    & 3& 3* \\ \hline 01000 & 3& 4* \\ \hline 010000 & 3& 5* \\ \hline
    0100000 & 3& 6* \\ \hline 001000 & 4& 5* \\ \hline 0010000 & 4& 6*
    \\ \hline
\end{tabular}  
\hspace{2cm}
\begin{tabular}{|l||c|c|} \hline A &\, $L_s $&\, $L_l$\\ \hline 010100
  & 5& $>50$ \\ \hline 0101000 & 6& $>50$ \\ \hline 010010 & 4& 10 \\
  \hline 0100100 & 4& 14 \\ \hline 0100010 & 4& 7 \\ \hline

\end{tabular}  
\end{center}
\label{tab:regular}
\end{table}

\section{Equivalence of the random and planted ensembles}
\label{sec:pl}

The planted ensemble of graphs, which is the main subject of the
present paper, is created in the following way:
\begin{itemize}
\item[(i)] Make each of the $N$ variables occupied with probability $\chi$
  (\ref{marginal_reg}), call the number of occupied variables $N_1$.
\item[(ii)] Choose a degree sequence from the probability distribution
  $Q(l)$ in such a way that $KM=\sum_{i=1}^N l_i$.
\item[(iii)] For each constraint, and according to the probabilities
  $x_r$ (\ref{prob_1}), choose the number $r_a$ of
  occupied variables to which it is connected. Repeat until
  $\sum_{a=1}^M r_a = \sum_{i=1}^{N_1} l_i$. Here $i$ are the indexes
  of the occupied variables. If this condition cannot be achieved go
  back to step (i) and repeat it until the condition is achievable.
\item[(iv)] Now consider the $r_a$ legs going out of every constraint $a$, order them arbitrarily and index them by $i$ going from $1$ to $\sum_{a=1}^M r_a$. Consider $l_i$ legs going out from every occupied variable and index them. Choose a random permutation $\pi$ of $\sum_{a=1}^M r_a$ numbers, and connect the leg with index $i$ going out from occupied variables to the leg with index $\pi(i)$ going out from constraints. Do the same with the empty variables and the remaining $K-r_a$ legs going out from the constraints. Repeat until there are no double edges.
\end{itemize}
Note that there are several other models how to plant a solution
(e.g. choose exactly the integer value of $\chi N$ occupied variables
in the step (i)), we could have chosen any other which is equivalent
(for typical properties) to the above one in the $N\to \infty$ limit.

\begin{definition}
  \label{def:therm}
  Call a property of a large random graph drawn from a given random
  ensemble a {\it thermodynamic property} if and only if in the $N\to
  \infty$ limit the probability that this property holds is smaller than
  $1-\exp(-cN)$, where $c$ is some $\Theta(1)$ constant.
\end{definition}
In statistical physics of random systems it is often the case that
large deviations are exponentially rare and hence all usually
considered properties are thermodynamic in this sense. Without
attempting a rigorous proof, in statistical physics the following
examples are often assumed to be thermodynamic properties: The degree
distribution, the entropy density, the fraction of occupied variables
in a random satisfying assignment, the distance between two random
satisfying assignments, etc.  Properties that are not thermodynamic
are all those relying on the behavior of exponentially rare instances,
e.g. moments of some exponentially large quantities, as for instance
the number of satisfying assignments.

\begin{definition}
  \label{def:equ}
  Consider two ensembles of random graphs $A$ and $B$, we call the two ensembles {\em asymptotically equivalent}
  if and only if every property that is thermodynamic in ensemble $A$ is thermodynamic in ensemble $B$, and vice versa.
\end{definition}
\begin{definition}
  The planting is called {\em quiet} if the corresponding planted and
  random ensembles are asymptotically equivalent.
\end{definition}

Quiet planting intuitively means that if one is given a large random graph, one is not able to tell if that graph was drawn from the random or from the planted ensemble. This is because properties that are usually measured to distinguish the two graphs ensembles are thermodynamic (even if proving they are thermodynamic might be in general difficult). 

The close relation between the random and the planted ensemble was
explored in \cite{AchlioptasCoja-Oghlan08} for the random graph
coloring and bi-coloring problems, see Theorem 6 and Theorem 7 in the
appendix A of \cite{AchlioptasCoja-Oghlan08}. In statistical physics
the quiet planting for graph coloring was discussed in
\cite{KrzakalaZdeborova09}.

\begin{proposition}
\label{planting}
Denote ${\cal N}_G$ the number of satisfying assignments of an
instance $G$ drawn from the random ensemble. If $ \mathbb E[\log{({\cal
    N}_G+1)]}/N$ is a thermodynamic property, i.e.  \be \exists c>0:
\forall \epsilon>0 \lim_{N\to \infty} P(|\log {\cal N}_G - \mathbb
E[\log{(\cal N}_G+1)]| > \epsilon N) < e^{-cN} \label{therm} \ee and if
the annealed entropy is equal to the
quenched one, i.e.  \be \log{ \mathbb E({\cal N}_G)}= \mathbb E[\log{(\cal N}_G+1)] +o(N). \label{annealed}
\ee then the planted ensemble and the random ensemble are
asymptotically equivalent.
\end{proposition}
\begin{proof}[of Prop. \ref{planting}]
  In the random ensemble we are drawing graphs uniformly from all
  the graphs with a given degree distribution. In the planted ensemble we
  are drawing from the same set of graphs but with probability proportional to
  ${\cal N}_G$. Since (\ref{therm}) and (\ref{annealed}) hold by
  assumption for the random ensemble, relation (\ref{therm}) holds for some $c'$ also for the planted ensemble. Indeed, if (\ref{therm}) holds and if there would be larger than exponentially small probability that planting draws instances with $|\log {\cal N}_G - \mathbb E[\log{(\cal N}_G+1)]| > \epsilon N$ then (\ref{annealed}) could not hold. For every other thermodynamic property, i.e. such that large deviations are exponentially rare, the the same argument applies.
\end{proof}

Statements equivalent to Prop.~\ref{planting} first appeared in
eq. (4) and Theorem 6 in \cite{AchlioptasCoja-Oghlan08}. Note that
when the concentration condition (\ref{annealed}) can be proven in a
stronger form, then the condition on the exponentially rare large
deviation can be weakened.

In statistical physics, using the cavity method arguments, the
condition (\ref{annealed}) can be evaluated. As we state at the end of sec.~\ref{sec:def}, when the Bethe entropy is asymptotically exact then the factorization and the equality of the quenched and annealed entropies are equivalent. Hence quiet planting is possible in all the factorized problems as long as the Bethe entropy is asymptotically exact. In the non-locked factorized problems, such as the random graph coloring, condition (\ref{annealed}) ceases to be true strictly before
the satisfiability threshold, as discussed in \cite{KrzakalaZdeborova09}. 

In the next section we argue that in the
factorized locked problems (\ref{annealed}) holds up to the
satisfiability threshold, and hence the planted and the random
ensembles are asymptotically equivalent for the factorized locked
problems in the whole range of parameters corresponding to the satisfiable phase on the random ensemble.

\subsection{Second moment argument}

Relation (\ref{annealed}) is in general rather hard to prove
rigorously. Achlioptas and Coja-Oghlan \cite{AchlioptasCoja-Oghlan08}
used instead a stronger condition $\mathbb E({\cal N}_G^2)<C\,
[\mathbb E({\cal N}_G)]^2$ which they proved for the coloring and
the bi-coloring of factor-graphs problem for sufficiently sparse graphs.

For the factorized locked problems we conjecture that the relation
$\mathbb E({\cal N}_G^2)<C\, [\mathbb E({\cal N}_G)]^2$ holds in all the
factorized locked problems on the purely random ensemble as long as
$\overline l \le l_s$.

The first and second moment of the number of solutions in the
occupation problems has been computed for a general degree
distribution in \cite{ZdeborovaMezard08b}.  Based on numerical results
it has been also argued non-rigorously in \cite{ZdeborovaMezard08b} that the above
conjecture holds in the balanced locked problems. Here we illustrate
that it also holds in the $R$-in-$K$ SAT on random $L$-regular graphs
for $L<l_s$.  The first moment entropy, defined by (\ref{eq:ann}), is in the
$R$-in-$K$ SAT on random $L$-regular graphs given by
eq. (\ref{ent_RinK}). The second moment entropy $s_{\rm
  2nd}=\lim_{N\to \infty}\log{\mathbb E({\cal N}_G^2)}/N$ is given by
$s_{\rm 2dn}= {\rm max}_t s_{\rm 2nd}(t)$ where
\cite{ZdeborovaMezard08b} \be s_{\rm 2nd}(t)= \frac{L}{K}\log{\left\{
    K!\sum_{s=0}^R \frac{ \left[ \left(\frac{tR}{K}\right)^{s}
        \left[\frac{(1-t)R}{K}\right]^{2(R-s)}
        \left[1+\frac{(t-2)R}{K}\right]^{K-2R+s}
      \right]^{1-\frac{1}{L}}}{(R-s)!\, (R-s)!\, s! \, (K-2R-s)!}
  \right\}}\, .\label{2nd} \ee The interpretation of the parameter
$0\le t\le 1$ follows from expression \be \mathbb{E}({\cal N}_G^2) =
\sum_{\sigma_1,\sigma_2} P(\sigma_1\, SAT, \sigma_2 \, SAT) \, ,\ee
where $\sigma_1$ and $\sigma_2$ are configurations and $P(\cdot)$ is a
probability over the graph ensemble. The parameter $t$ in (\ref{2nd})
is then the number of sites occupied in both $\sigma_1$ and $\sigma_2$
divided by number of sites occupied in one of the solutions, $RN/K$.
We remind that in the $R$-in-$K$ SAT the satisfiability threshold is
given by cancellation of the entropy (\ref{ent_RinK}) \be l_s= \left[
  1- \frac{\log{{K\choose R}}}{K H\left(\frac{R}{K}\right)}
\right]^{-1}\, . \ee

\begin{figure}[!ht]
\begin{center}
\includegraphics[width=0.49\linewidth]{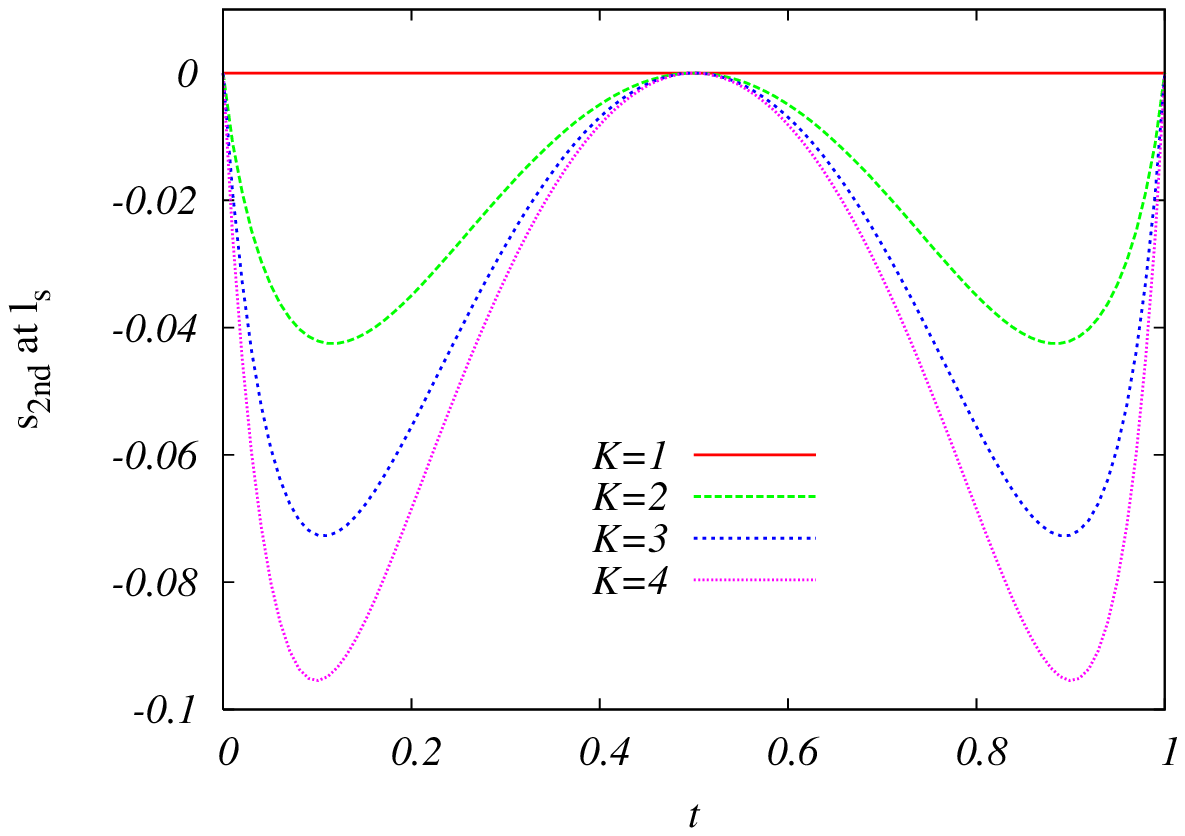}
\includegraphics[width=0.49\linewidth]{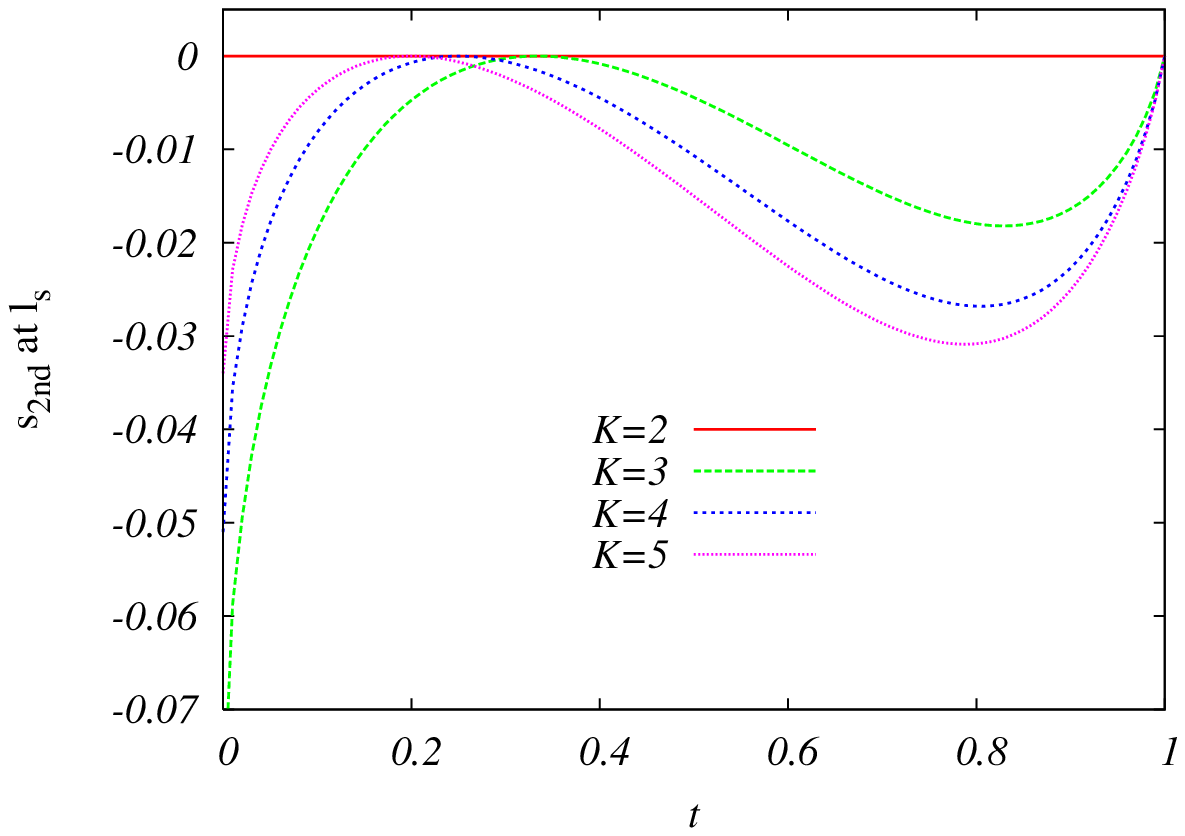}
\end{center}
\caption{\label{kin2k_2nd} The second moment entropy function $s_{\rm
    2nd}(t)$ (\ref{2nd}) at $l_s$ for several values of $K$, in the $K$-in-$2K$ SAT problem on the left, and $1$-in-$K$ SAT on the right.}
\end{figure}

As $s_{\rm 2nd}$ is a maximum of a function of a single variable $t$,
we plot $s_{\rm 2nd}(t)$ at $L=l_s$ in Fig.~\ref{kin2k_2nd}. Evaluation of the polynomial function (\ref{2nd}) for many values of $R$ and $K$ in Mathematica shows that for $L<l_s$ we have $2s_{\rm ann}=s_{\rm 2nd}\ge 0$, and for $L>l_s$ we have $s_{\rm ann}=s_{\rm 2nd}\le 0$.

We also investigated numerically general formulas for the second moment presented in \cite{ZdeborovaMezard08b} and concluded that 
$2s_{\rm ann}=s_{\rm 2nd}$ for $l<l_s$, and $s_{\rm ann}=s_{\rm 2nd}\le 0$ for $l>l_s$, holds also for all the other locked factorized problems. 

\subsection{Satisfiable factorized locked problems equivalent to the planted ones}

We also conjecture that in the factorized locked models the planted ensemble is asymptotically equivalent to the ensemble of satisfiable instances in the whole region of $\overline l$. 

For a general (non necessarily locked) constraint satisfaction problem
the space of satisfying assignments is separated into clusters. We
define the entropy $s$ of a cluster as the logarithm of the number of
assignments that belong to this cluster. We also define the complexity
$\Sigma(s)$ as the logarithm of the number of clusters of a given
entropy $s$. The function $\Sigma(s)$ is well defined even in the
unsatisfiable region: when $\Sigma(s)<0$ it then corresponds to the
large deviation function for the existence of a cluster of a given
size \cite{Rivoire05}. It was argued in \cite{KrzakalaZdeborova09}
that the cluster containing the planted configuration has a size $s^*$
such that $s^*={\rm argmax} [\Sigma(s)+s]$. On the other hand from the
large deviation interpretation of the $\Sigma(s)$ function, most of the
rare satisfiable instances in the unsatisfiable region will have one
cluster of size $s'={\rm argmax}\, \Sigma(s)\le s^*$.

In the locked problems, all clusters contain a single satisfying
assignments, hence $s=0$ for all clusters. Keeping in mind the large
deviation interpretation of the complexity $\Sigma$ \cite{Rivoire05},
the rare satisfiable instances have a single solution and should be
asymptotically equivalent to planted instances. And hence the
satisfiable and planted ensembles are asymptotically equivalent in the
whole range of $\overline l$ in the factorized locked problems.

\section{Single solution instances}
\label{sec:sin}

As discussed in the introduction, it is of practical importance to be
able to create hard instances which have a single solution with a large
probability. Based on the heuristic cavity method results of
\cite{KrzakalaZdeborova09} we conjecture that in the region $\overline
l > l_s$ with high probability there is a single solution on large
planted instances of the factorized locked problems (or a couple in
case of balanced problems). In this section we prove (assuming the properties of the 1st and 2nd moment from the previous section) this statement
for the $R$-in-$K$ SAT on random regular graphs. We believe that the 
generalization of the proof is possible also for the other factorized
locked problems.

\begin{figure}[!ht]
\begin{center}
\includegraphics[width=0.49\linewidth]{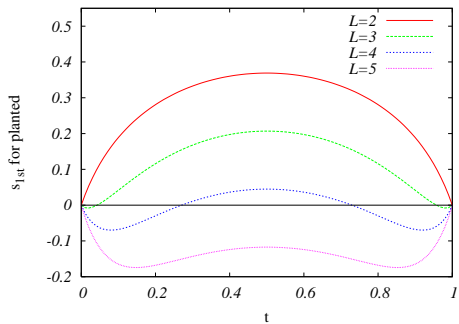}
\end{center}
\caption{\label{kin2k_pl} The first moment entropy in the $4$-in-$8$ SAT on
  $L$ regular planted ensemble.}
\end{figure}

First note that the first moment in the planted $R$-in-$K$ SAT $s_{\rm
  ann,pl}={\rm max}_t s_{\rm
  ann,pl}(t)$ is related in a simple way to the first and second moment in the purely
random ensemble. It holds for the entropies \be s_{\rm
  ann,pl}(t)=s_{\rm 2nd}(t)-s_{\rm ann}\, . \ee See an example of the
function $s_{\rm ann,pl}(t)$ in Fig.~\ref{kin2k_pl}.

From the previous section it follows that for $L>l_s$ the first
moment entropy in the planted ensemble is a negative function for all
$0<t<1$. The parameter $t$ is in the planted ensemble interpreted as
the distance from the planted solution. Therefore, for $L>l_s$ there
are no solutions at an extensive (i.e. $\Theta(N)$) distance from the
planted solution (except the solution at distance one in the balanced
problems).

\begin{theorem}
\label{expander_1}
Consider a large instance of the $R$-in-$K$ SAT problem drawn from the
planted ensemble, the degree of variables be $L>2$. Then there exists an
$\epsilon>0$ such that with high probability there is no solution at
distance smaller than $\epsilon N$ from the planted solution
\end{theorem}
We will use the expander properties of regular bipartite graphs. The
following theorem is well known in the theory of expanders
\cite{SipserSplielman96}.
\begin{theorem}
\label{expander}[Sipser and Spielman \cite{SipserSplielman96}]
Consider a random factor-graph with degree of variables $L$ and degree
of constraints $K$. Then, for any $\delta<L-1$, there exists a
constant $\epsilon>0$, such that with high probability for every set
of $\tilde N\le \epsilon N$ variables the number of neighboring
constraints is larger than $\delta \tilde N$. In other words the
factor graph is a $(\epsilon,\delta)$ expander.
\end{theorem}
\begin{proof}[of Theorem \ref{expander_1}] Let us prove the statement
  by contradiction. Suppose that as $N\to \infty$ for every
  $\epsilon>0$ there is a solution at distance smaller that $\epsilon
  N$ from the planted solution. Denote the distance between the
  planted and this nearby solution $N_1=\epsilon'N$.  Now consider
  the factor-graph and the planted solution, $N_1$ of variables have
  to be changed to reach the nearby solution.  Since $\epsilon'$ can
  be arbitrarily small Theorem \ref{expander} implies that there is at
  least $\delta N_1$ constraints in which at least one variable has
  been changed. The property defining a locked constraint is that if a
  variable is changed then at least one other has to be changed in order to
  satisfy the constraint again. Hence each of the at least $\delta
  N_1$ constraints have to be connected by at least two edges to the $N_1$ changed
  variables. There is hence at least $2 \delta N_1$ edges connected to
  changed variables. The total degree of changed variables is $L N_1$,
  hence $L N_1>2 \delta N_1$. But as $\delta$ can be as near to $L-1$
  as we wish this inequality cannot hold and we hence reached a
  contradiction. This proofs that there exists $\epsilon>0$ such that
  with high probability there is no solution at distance smaller than
  $\epsilon N$ from the planted one.
\end{proof}

Properties of the first moment in the planted ensemble together
with Theorem \ref{expander_1} imply that in the planted $R$-in-$K$ SAT
on random regular graphs there is almost surely a single solution (or
a pair of solutions for $R=K/2$).

\begin{figure}[!ht] 
\includegraphics[width=0.495\linewidth]{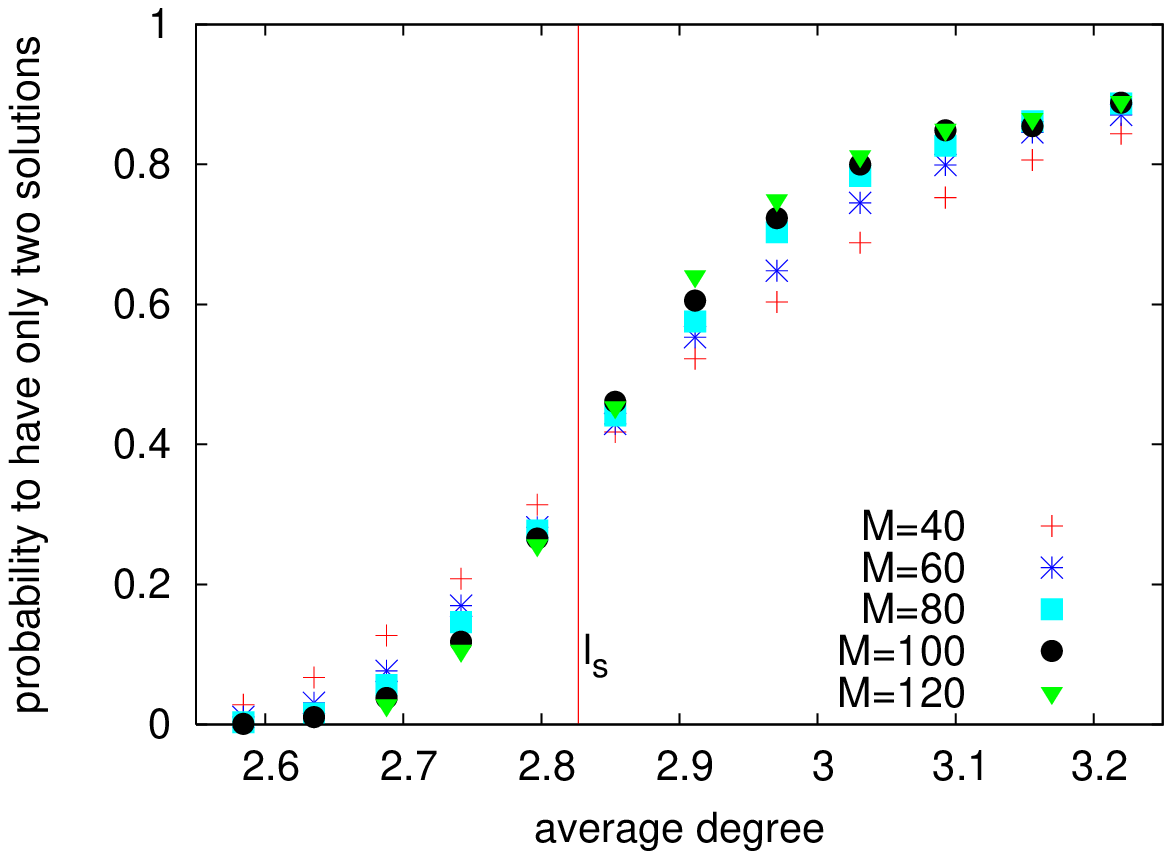}
\includegraphics[width=0.495\linewidth]{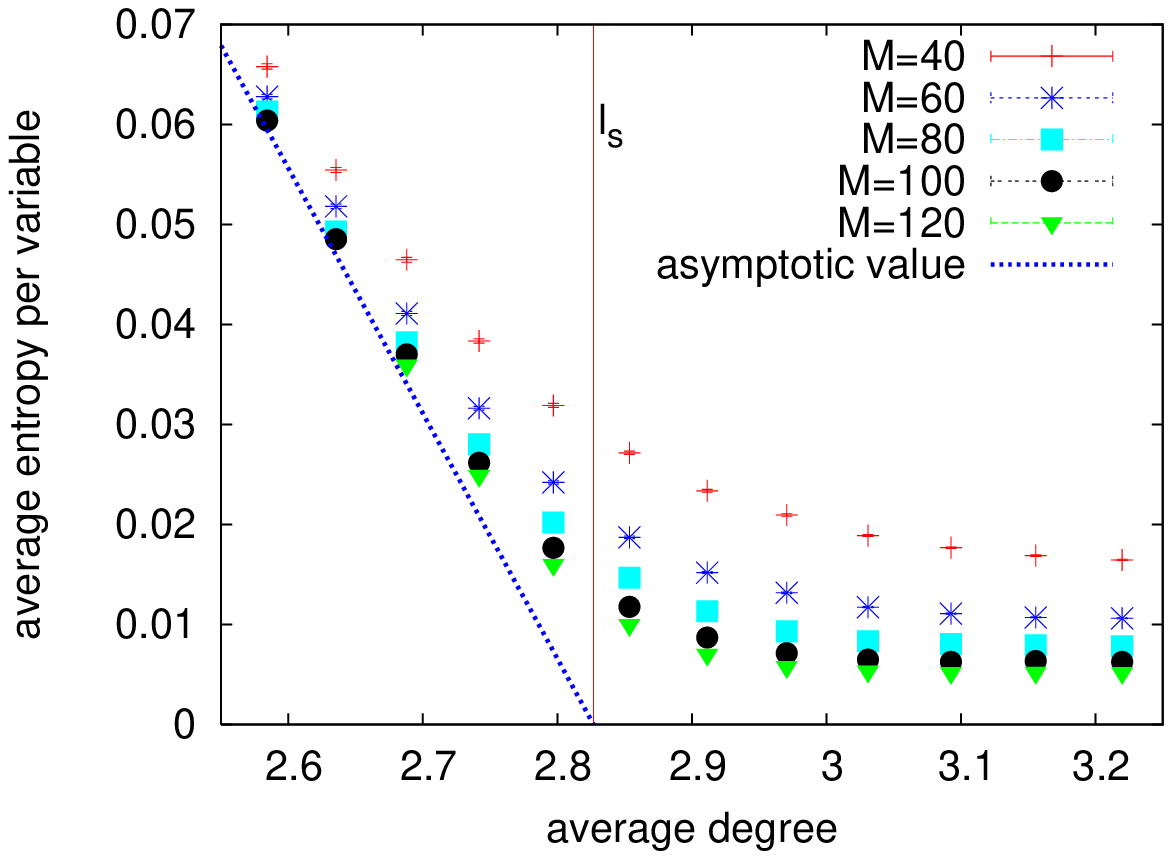}
\caption{\label{pdep} Left: Probability (over $5000$ instances) that
  there is a single pair of solutions in the $2$-in-$4$ SAT as a function
  of the average degree and the size of the graph. Right: Data are the average
  entropy density (logarithm of the number of solutions per variables)
  of the instances. The line represents the entropy density in the $N\to \infty$ limit, eq. (\ref{ent_reg}). The data are obtained with the {\tt relsat}
  algorithm \cite{BayardoPehousek00}. In both parts we marked the threshold $l_s=2.827$.}
\end{figure}  

\section{Average computational hardness}
\label{sec:alg}

One of the most interesting aspects of the study of random constraint
satisfaction problems is the average computational hardness of a given
ensemble. This has been discussed extensively in both the computer
science and the physics literature, in particular for the
K-satisfiability and coloring problems. It has been shown empirically
that the hardest instances lie very near to the satisfiability
threshold $l_s$, and an easy-hard-easy pattern is often described
\cite{CheesemanKanefsky91,MitchellSelman92}.  Later works focused on
predicting up to which connectivity polynomial algorithms are able to
find solutions, see
e.g. \cite{MezardParisi02,ZdeborovaKrzakala07,SeitzAlava05}. Instances
with a very large density of constraints are typically
unsatisfiable. In some problems, e.g. K-satisfiability, no on average polynomial
algorithms are known to show unsatisfiability for arbitrary large but
constant density of constraints \cite{ChvatalSzemer88}. In other, more
constraint, problems unit clause propagation based schemes were shown
to be efficient \cite{AchlioptasChtcherba01}.  In the planted
instances, which are always satisfiable, it is known that for
sufficiently large density of constraints solutions can be found in
polynomial time, see
e.g. \cite{KrivelevichVilenchik06,CojaOghlanMossel07}. The situation
in the planted factorized locked problems is very interesting: on top
of the easy low and high constraint density phases we show that there
also exists an intermediate hard phase, and we locate both the
boundary thresholds.

We argued that in the satisfiable phase $\overline l <l_s$ the planted
and random ensembles are asymptotically equivalent, this includes the
average behavior of algorithms. It was argued in
\cite{ZdeborovaMezard08,ZdeborovaMezard08b} that for average degree
$\overline l < l_d$ the locked problems are algorithmically easy
whereas for $l_d< \overline l < l_s$ they are on average hard.

The second hard-easy transition is particular to the planted ensemble
and happens in the unsatisfiable phase $\overline l >l_s$.  We will study the behavior of
the BP equations initialized randomly to locate this transition.

\subsection{The spinodal point}

By definition of the factorized locked problems the belief propagation
equations (\ref{BP1}) initialized randomly converge to a uniform fixed
point. But as the average degree is growing this ceases to be true. In
the problems that we are studying here, there actually exists a
critical average degree $l_l$ beyond which belief propagation
converges spontaneously towards the planted solution. This yields a
clear hard-easy transition in the algorithmic complexity. In
statistical physics terms this threshold $l_l$ corresponds to a
spinodal point of the liquid state \cite{KrzakalaZdeborova09}. The
spinodal point also corresponds to the Kesten-Stigum bound
\cite{KestenStigum66,KestenStigum66b}, and to the robust
reconstruction threshold on trees \cite{JansonMossel04}.  This is yet
another important connection between the reconstruction problem and
the planted ensemble.

In order to compute the spinodal point let us first define matrix
$z(s'|s)$. Consider a variable and one of its neighbors, $z(s'|s)$ is
then the probability that in the planted configuration the variable
was assigned $s'$ given that its neighbors was $s$. In the terms on
reconstruction on trees $z(s'|s)$ is the probability that in the
broadcasting a variable was assigned $s'$ given its parent was $s$.
Components of $z(s'|s)$ can be computed as \bea
z(0|0)=\sum_{r=0}^K \left(1-\frac{r}{K-1}\right) y_r(0)\, , \quad \quad z(1|0)=1-z(0|0)\, , \\
z(1|1)=\sum_{r=0}^K \frac{r-1}{K-1} y_r(1) \, , \quad \quad
z(0|1)=1-z(1|1) \, .\eea 
where
 \be y_r(0)=\frac{(K-r)\,
  x_r}{\sum_{t=0}^K (K-t)\, x_t }\, , \quad \quad y_r(1)=\frac{r\,
  x_r}{\sum_{t=0}^K t\, x_t }\, , \label{prob_ex} \ee
where $x_r$ is given by (\ref{prob_1}) or (\ref{prob_2}). Explicit formulas for the regular problems are 
\bea
z(0|0)  =  \frac{\sum_{r=0}^{K-2}  {K-2 \choose r} \delta_{A(r),1} \psi_1^{r(L-1)} \psi_0^{(K-r-1)(L-1)} }{\sum_{r=0}^{K-1}  {K-1 \choose r} \delta_{A(r),1} \psi_1^{r(L-1)} \psi_0^{(K-r-1)(L-1)}} \, ,\\
z(1|1) = \frac{\sum_{r=2}^{K} {K-2 \choose r-2} \delta_{A(r),1}
  \psi_1^{(r-1)(L-1)} \psi_0^{(K-r)(L-1)} }{\sum_{r=1}^{K} {K-1
    \choose r-1} \delta_{A(r),1} \psi_1^{(r-1)(L-1)}
  \psi_0^{(K-r)(L-1)}} \, .
\eea

The first eigenvalue of this matrix is equal to one, and is associated
with a trivial homogeneous eigenvector. The second eigenvalue of the
matrix $z$ is given by \be \lambda = z(0|0)+z(1|1)-1\, .
\label{lambda}
\ee

A well-known property of the reconstruction on a tree is that
reconstruction is always possible beyond the so called Kesten-Stigum
(KS) threshold \cite{KestenStigum66,KestenStigum66b}. In our notation
the KS condition says that if $(L-1)(K-1)\lambda^2>1$ then the
reconstruction is possible, i.e., the leaves asymptotically contain
some information about the value sent by the root. In statistical physics the Kesten-Stigum condition is equivalent to the de Almeida-Thouless instability of the paramagnetic phase towards a spin-glass phase \cite{AlmeidaThouless78,MezardMontanari06,KrzakalaMontanari06}, that is for
$(L-1)(K-1)\lambda^2>1$ the belief propagation equations (\ref{BP1}) do
not converge. This can be seen from the fact that \be \lambda =
\frac{\partial \psi^{a\to i}_1 }{\partial \psi^{b\to
    j}_1}\, , \label{deriv} \ee where $j\in \partial a \setminus i$, and
$b\in \partial j \setminus a$.

The eigenvalue $\lambda$ and the condition for reconstructibility
$(L-1)(K-1)\lambda^2>1$ also appear in the problem of robust
reconstruction on trees \cite{JansonMossel04}. In the problem of
robust reconstruction it is required that even if an arbitrary large
fraction of the values on the leaves is erased there is still
information about the root left.

The analysis of the instability of the uniform BP fixed point towards
the planted solution then goes as follows. Consider a part of the
factor-graph as depicted in Fig.~\ref{fig_cav}. Denote the values of
the messages in the uniform fixed BP fixed point by
over-bars. Consider the incoming message to be perturbed from the
uniform value as \be \left( \begin{array}{c}
    \psi^{b\to j}_1 = \overline{\psi_1}+ \epsilon \\
    \psi^{b\to j}_0 = \overline{\psi_0}- \epsilon \\\end{array}
\right) \, .\ee Note that $\epsilon$ can be both negative or positive. The
equation (\ref{deriv}) then implies that the outgoing message will be
\be \left(
\begin{array}{c}
    \psi^{a\to i}_1 = \overline{\psi_1}+ \lambda \epsilon \\
    \psi^{a\to i}_0 = \overline{\psi_0}- \lambda \epsilon \\
\end{array}
\right) \, .\ee In other words, any infinitesimal noise in one of the
incoming message is multiplied by $\lambda$ in the recursion.

We call the perturbation of the incoming message $\epsilon_+$ if $j$
was occupied in the planted configuration, and $\epsilon_-$ otherwise.
If the variable $i$ was planted in the occupied state, then $j$ was
planted occupied with probability $z(1|1)$, and empty with probability
$z(0|1)$. Similarly, if the variable $i$ was planted in the empty
state, then $j$ was planted empty with probability $z(0|0)$ and
occupied with probability $z(1|0)$. Thus the evolution of the
perturbation is governed by the equation: \be \left( \begin{array}{c}
    \epsilon^{a\to i}_+ \\
    \epsilon^{a\to i}_- \\\end{array} \right) = \lambda
\left( \begin{array}{cc}
    z(1|1) & z(0|1) \\
    z(1|0) & z(0|0) \end{array} \right) \left( \begin{array}{c}
    \epsilon^{b\to j}_+ \\
    \epsilon^{b\to j}_- \\\end{array} \right)\, . \ee

Moreover there are $(K-1)(L-1)$ possible incoming messages in the
regular graph, thus the criterion $(K-1)(L-1) \lambda^2=1$. If
$(K-1)(L-1) \lambda^2<1$ then the perturbation decreases and we find
only the uniform BP fixed point, if on the contrary $(K-1)(L-1)
\lambda^2>1$ the uniform BP fixed point is unstable and a
perturbation towards the planted configuration amplifies
exponentially.

As the planted configuration corresponds to a stable BP fixed
point\footnote{Note that in the above calculation we considered
 the stability around the uniform BP fixed point, if we consider the BP
  fixed point corresponding to the planted solution the perturbation
  does not amplify.} the BP iterations converge instead to the planted
solution. Fig.~\ref{BP_fig} confirms that this is true even on rather
small graphs. On the balanced locked problems, where we are not
restricted to regular graphs, the correct condition is $(K-1)
\gamma \lambda^2=1$, where $\gamma$ is the mean of the excess degree
distribution $q(l)= (l+1) Q(l+1)/\overline l$. The spinodal point
$l_l$, see Tabs.~\ref{tab:balanced},\ref{tab:regular}, is then defined
by \be (K-1)(l_l-1) \lambda^2=1 \ee for the regular graphs, and \be
(K-1) \lambda^2= \frac{1-e^{-c_l}}{c_l} \ee for the truncated
Poissonian distribution.

The existence of this spinodal point, together with the conjecture
about equivalence between the planted ensemble and the ensemble of
satisfiable instances from the random ensemble, Sec.~\ref{sec:pl},
implies that for $\overline l > l_l$ it is easy to recognize almost
all satisfiable instances of the locked problems. Similar conclusions,
without a sharp threshold, were established for the coloring and
satisfiability problems in
\cite{Coja-OghlanKrivelevich07,AltarelliMonasson07}.

\subsection{Belief propagation as a solver}

Belief propagation reinforcement is a good solver in the region
$\overline l<l_d$ as shown empirically in
\cite{ZdeborovaMezard08,ZdeborovaMezard08b} in the random ensemble. Since
the two ensembles are equivalent in that region, nothing changes for the planted ensemble. We have indeed verified this numerically.

\begin{figure}[!ht] 
\includegraphics[width=0.495\linewidth]{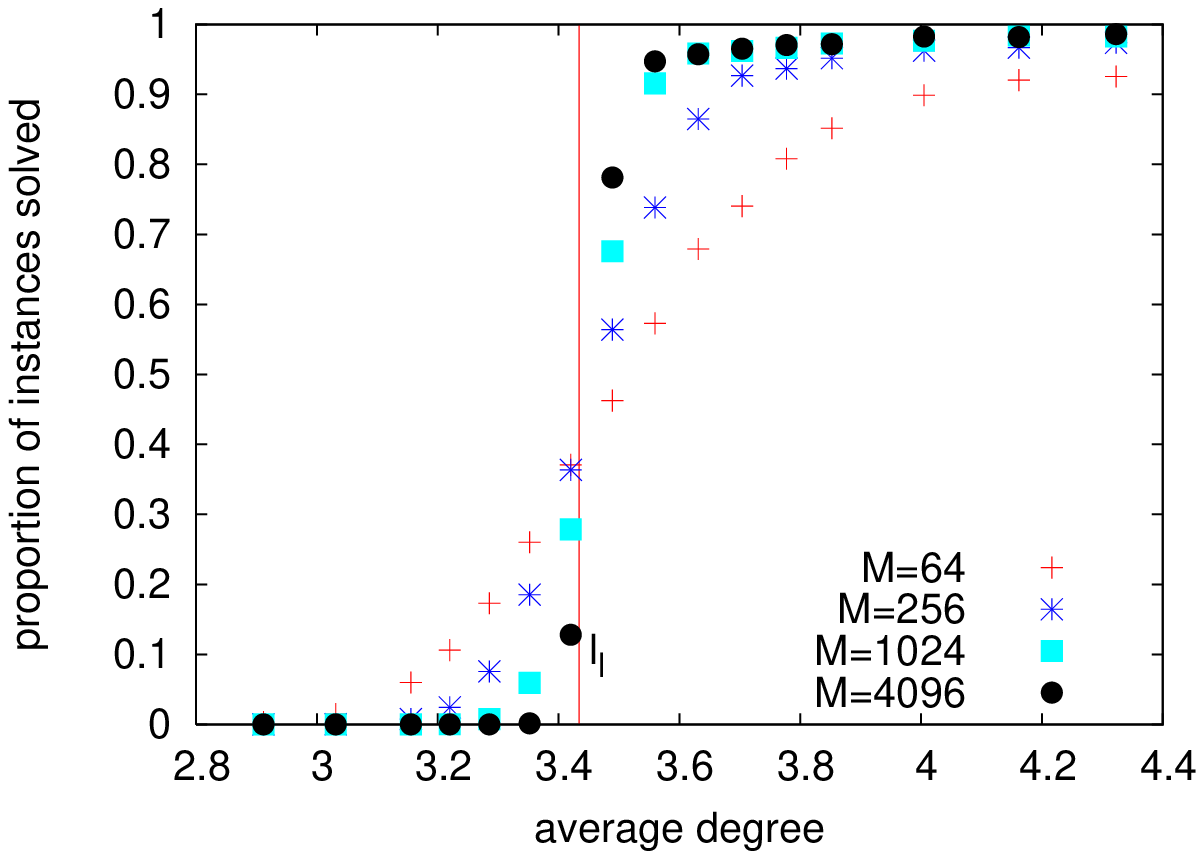}
\includegraphics[width=0.495\linewidth]{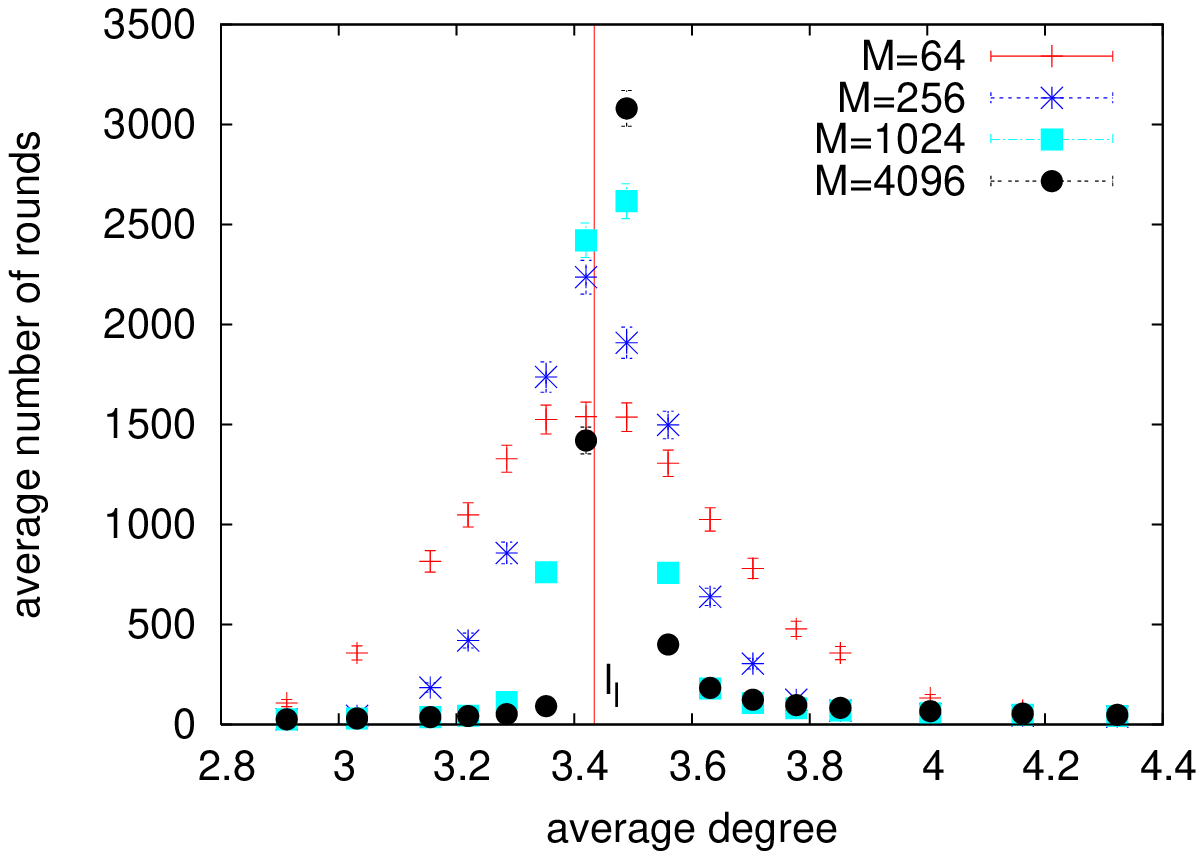}
\caption{\label{BP_fig} Belief propagation on the $2$-in-$4$ SAT problem. Left: Probability that the belief propagation
  algorithm finds the planted configuration when initialized randomly
  plotted as a function of the average degree for several system
  sizes. Right: The convergence time dependence on the average
  degree. In both cases, we have stopped the BP iterations when the average change per
  message was less than $2.10^{-3}$. In both parts we marked the spinodal threshold $l_l=3.434$.}
\end{figure}

Based on the above arguments, belief propagation equations
converge to the uniform fixed point for $\overline l< l_l$ and
directly to the planted solution for $\overline l>l_l$. In order to
verify that on finite size instances, we have performed the following
numerical experiment: we have generated many planted instances for
different sizes and average degrees ($5000$ instances for each set of
parameters). We then iterated the BP equations (\ref{BP1}) starting
from random initial conditions. For numerical stability reasons we used
dumping in the iterations, i.e. each time we computed a new message we
kept one half of the sum of the new and old message. As a convergence
criterion we used that the messages should not change more that
$2.10^{-3}$ per message (we checked that a smaller criterion does not
change the quality of results, and only slows down the computation).
This way every iteration converged either to a configuration where the
bias of each variable pointed towards the planted solution (or to its
negation) or to a point very near to the uniform fixed
point. Fig.~\ref{BP_fig} shows in what fraction of the runs we were
able to find the planted solution and in particular it confirms that
for $\overline l >l_l$ it is easy to find it in linear time. On the
right of the same figure we plot the average convergence time (given
the criterion $2.10^{-3}$ per message). We see that around the
spinodal point $l_l$ the convergence time diverges from both the sides
(slightly faster from the large degree side).

\section{Conclusions and perspectives} 
\label{con}

In this work we have studied a class of constraint satisfaction
problems on a planted ensemble. The solution is planted in a {\it
  quiet} way, i.e. the planted configuration is one of the typical
solutions of the resulting instance. So far we know how to realize such
plantings only on the factorized problems. We describe several
connections between this quiet planting and the problem of
reconstruction on trees.

We study the locked problems because of the simple structure on the
space of their solutions --- solutions are isolated points instead of
clusters. This property makes the locked problems, however, very hard
algorithmically. We focused on the class of occupation locked problems
in this manuscript, all our results generalize easily to any factorized locked
problem, on non-binary variables for example.

On non-locked but factorized problems, as e.g. graph coloring,
the concept of quiet planting stays valid \cite{KrzakalaZdeborova09},
however, the random and planted ensembles are not equivalent up to the
satisfiability threshold. Moreover, in the unsatisfiable phase the planted
ensemble has exponentially many solutions, instead of a single one as
is the case in the locked problems. The non-locked problems are also
much less friendly for first and second moment considerations. The
phase diagram of the locked but non-factorized problem will not be
very different from the one presented here. However, the thresholds
will be different in the planted and random ensembles and the two
ensembles are not equivalent.

One of the most important results of our work is the location of the
algorithmically hard region, between $l_d \le \overline l \le l_l$, in
the problems under investigation. It would be in particular
interesting to design an algorithm which would provably find solutions
in the region $\overline l > l_l$, as we have only heuristic and
numerical arguments. This is also challenging in the non-locked
problems, as e.g. graph coloring, where we predicted the planted
Poisson ensemble to be easy above $l_l=(q-1)^2$ (on planted regular
graphs $L_l=(q-1)^2+1$), where $q$ is the number of colors. Results
establishing that the planted ensemble on coloring is easy above
$Cq^2$, where $C$ is some constant quite larger that one, are already
known \cite{KrivelevichVilenchik06,CojaOghlanMossel07}.

Finally, another consequence of our work worth discussing is that we
know how to generate unique satisfying assignment instances - both in
the hard and easy regions. Such instances are often used for
evaluating the performance of the quantum annealing algorithm, but so
far they have been generated with an exponential
cost from an ensemble with unknown classical average computational
complexity \cite{YoungKnysh08,FarhiGoldstone01}. In our opinion, these
works should be repeated on instances of the locked problems. We
conjecture that in the classically hard region also the quantum
annealing will be exponential (this is because we anticipate a first
order phase transition in the transverse magnetic field, as in \cite{JoergKrzakala08}).

\section*{Acknowledgments} 
We thank to Dimitris Achlioptas, Amin Coja-Oghlan, Matti Jarvisalo, Andrea Montanari, Cris Moore and Guilhem Semerjian for fruitful discussions and suggestions.

\bibliography{myentries}

\end{document}